\documentclass{PoS}
\let\ifpdf\relax
\usepackage{rotating}
\usepackage{multirow}
\usepackage[titletoc,title]{appendix}
\usepackage{natbib}
\bibpunct{(}{)}{;}{a}{}{,}

\title{Revealing the Physics and Evolution of Galaxies and Galaxy Clusters with SKA Continuum Surveys\footnote{This chapter is presented on behalf of the {\it SKA Continuum Science Working Group}, and builds upon work done in consultation with a wider expert community. A comprehensive discussion of the various science topics summarized here is presented in a number of dedicated chapters.}}

\ShortTitle{Continuum Surveys: Galaxies and Galaxy Clusters}

\author{
\speaker{Isabella Prandoni}$^1$ \&
Nicholas Seymour$^2$
\\ 
$^1$INAF - IRA, Via P. Gobetti 101, 40129 Bologna (Italy)\\
$^2$12Curtin Institute of Radio Astronomy, Curtin University, GPO Box U1987, Perth WA 6845, Australia\\
\\
E-mail: \email{prandoni@ira.inaf.it , nick.seymour@curtin.edu.au}
}

\abstract{In this chapter we provide an overview of the science enabled by radio continuum surveys in the SKA era, focusing on galaxy/galaxy cluster physics and evolution studies, and other relevant continuum science in the $>2020$ scientific framework.  
We outline a number of {\it reference} radio-continuum surveys for SKA1 that can address such topics, and 
comprehensively discuss the most critical science requirements that we have identified.
We highlight what should be achieved by SKA1, to guarantee a major leap forwards with respect to the pre-SKA era, considering  the science advances expected in the coming years with existing and upcoming telescopes (JVLA, LOFAR, eMERLIN, and the three SKA precursors: MWA, ASKAP and MeerKAT). In this exercise we take in due account also the other waveband facilities coming online at the same time (e.g. {\it Euclid}, LSST, etc.), which tackle overlapping scientific goals, but in a different manner. In this respect particular attention has been payed to ensure that the proposed {\it reference} surveys are able to exploit the existing  synergies with such facilities, so as to generate strong involvement from all astronomical communities, and leave a lasting legacy value.
It is clear that a certain degree of freedom is allowed to some of the observational parameters. We believe it is very important to best fine-tune such parameters taking into proper account existing commensalities with SKA1 surveys addressing other science areas (HI galaxy  science, magnetism, cosmology). 
}

\FullConference{
Advancing Astrophysics with the Square Kilometre Array\\
June 8-13, 2014\\
Giardini Naxos, Italy}

\newcommand{\skipthis}[1]{}

\begin{document}

\section{Preamble}

\noindent 
Radio continuum surveys are a common tool  that can be exploited to address several scientific areas like galaxy evolution, magnetism, transients, cosmology etc. Here we present those science cases that are the focus of the {\it SKA Continuum Science Working Group (WG)}, primarily in the field of galaxy/galaxy clusters' physics and evolution. \\
The WG covers a wide range of science areas and represents a diverse cross-section of the astronomy community, demonstrating that a wide census of galactic and extra-galactic research areas can be addressed by the SKA, provided that key continuum science requirements are met.  
Such diversity is comprehensively illustrated in the collection of {\it Continuum Science} chapters presented in this Volume. \\
The top science priorities/drivers for continuum SKA1 surveys are based on the {\it Continuum Science} chapters, and have been identified from the WG after a long process that involved the wider community through a dedicated {\it Science Assessment Workshop} (Sept. 2013), the establishment of an {\it ad-hoc} SKA1-LOW {\it tiger team} (October-March 2014), and a breakout session on radio--continuum surveys at the AASKA14 Conference (June 2014). \\
The WG scientific discussion took into account (a) the expected scientific framework in years 2020-2025; (b) the fact that SKA1 should go a major step further with respect to its pathfinders (e.g. JVLA, LOFAR, eMERLIN, etc.) and precursors (Meerkat, MWA and ASKAP); (c) other facilities coming online at the same time (i.e. the planned facilities in the other wavebands, e.g. {\it Euclid}, LSST, etc.) which tackle overlapping scientific goals, but in a different manner. \\
With all this in mind, four {\it reference} radio--continuum surveys were defined, that are considered to have relevant scientific impact in the SKA1 era.
Such surveys are designed and optimized so as to respond to specific major scientific drivers, in key research areas like galaxy evolution and cosmology, but will be able to address a much broader range of science cases. As with most modern radio continuum surveys, these are naturally of the form of a nested `wedding cake' going from all-sky/wide to deep/ultra-deep fields. We note that the surveys presented here have been made compliant with the outcome of the SKA1 rebaselining process. 

\section{Top Priority Continuum Science Cases}
\noindent 
In this section we briefly summarize the top priority science cases in the field of galaxy/galaxy cluster physics and evolution, that we use as the main drivers for our {\it reference} surveys. For a more comprehensive discussion of such science cases, we refer to the  dedicated chapters in the present Volume, as detailed below.   It s clear that radio--continuum surveys can serve other relevant research fields. Key science drivers for magnetism, transients, cosmology, etc. are summarized by
the relevant overview chapters in this Volume. 

\vspace{0.5cm}\noindent
{\bf $\diamond$ The Star Formation History of the Universe} 

\vspace{0.2cm}\noindent
Measuring the Star Formation History of the Universe (SFHU) is a key goal of modern astrophysics, as it allows us to determine the how, when and why of star formation across the whole of cosmic time.  Radio continuum observations provide a unique way of measuring the star formation rate (SFR) in galaxies, as the synchrotron emission from relativistic supernova particles 
and free-free emission from H{\sc II} regions are unaffected by dust obscuration (unlike optical/UV) and directly scale to SFR. 
The SKA can potentially play a major role in assessing the SFHU since its phase 1, as
the sensitivity and resolution of the SKA1 at Band 1 and/or 2 provide the means to 
reach much lower SFRs than confusion limited far-IR surveys. 

\vspace{0.2cm}\noindent
Observations at $\nu\geq 10\,$GHz (Band 5) will provide higher resolution imaging and 
probe higher rest-frame frequencies with increasing redshift, where free-free 
processes dominate. In doing so, Band 5 surveys can provide
a more direct, dust-unbiased measurement of the massive star formation rate 
by being highly sensitive to the number of ionizing photons that are produced.  Having observations at both Bands 1 (or 2) and 5 we can have insights on the ISM
properties at high redshifts, and compare RC (Radio Continuum) - SFR relations
independently obtained from thermal and non-thermal physical processes. In addition Band
5 will provide angular resolutions of the order of 0.05-0.1 arcsec, needed to map the
distribution of active star formation within galaxies, and morphologically separate out AGN
contributions to total energetics for unbiased measurements of star formation rates. This
type of imaging is extremely complementary to ALMA imaging of CO for resolved studies of
the Schmidt law at very early epochs.

\vspace{0.2cm}\noindent
Galaxy evolution is best addressed through a tiered survey strategy whereby enough volume is sampled at each cosmic epoch of interest in order to overcome sample variance and gain a representative view of the Universe, from voids to  clusters.

\vspace{0.2cm}\noindent
{\it For more details see \citeauthor{jarvis15,murphy15,mcalpine15,mancuso15,ciliegi15} (\citeyear{jarvis15}, this Volume).}
  
\vspace{0.5cm}\noindent
{\bf $\diamond$ The Role of Black Holes in Galaxy Evolution} 

\vspace{0.2cm}\noindent
Super Massive Black Hole (SMBH) accretion occurs in at least two different modes: (a) a radiatively efficient {\it fast} accretion mode, also known as {\it quasar mode}, typically associated with radio-quiet (RQ) active galactic nuclei (AGN), (b) a radiatively inefficient {\it slow} accretion mode, also known as {\it radio mode}, typically associated with radio-loud (RL) AGN. Each mode is associated with different ``feedback'' processes. AGN feedback processes have become a standard ingredient in semi-analytic models of galaxy evolution in order to reproduce observed galaxy properties, but a clear understanding of these complex processes and their true role in shaping galaxy evolution remains elusive: what is the  relative importance of the two feedback modes as a function of galaxy mass and epoch? Which role is played by the environment? What drives these processes? \\
Large-areas are particularly important for AGN feedback studied as a sufficicient volume has to be probed to include the full range of galaxy environments, given the very important role that large-scale environment can play in triggering radio-mode feedback. 

\vspace{0.2cm}\noindent
The discovery of an increasing fraction of RQ AGNs in the current deepest radio
fields means that future sensitive radio surveys can in principle probe the entire AGN population, and not only the tiny (10\%) RL AGN fraction. The advantage of radio over e.g. X-ray/optical studies is that radio emission is not affected by gas obscuration, so $\mu$Jy and sub-$\mu$Jy radio surveys are sensitive to all types of AGN, independently of obscuration and their orientation (i.e., Type 1s and Type 2s). 

\vspace{0.2cm}\noindent
SKA1 will push evolutionary studies into the realm of the RQ AGN regime, at the same time being sensitive to the onset and earliest evolution of the RL AGN phenomenon in the Universe. By achieving a much higher sensitivity over very large survey areas, SKA1 will be able to reach well into the epoch of formation of the earliest AGN ($z>7$). Besides the immediate interest for galaxy evolution, the successful detection of a RL AGN at very high redshifts would allow for the direct study of neutral Hydrogen in the Epoch of Reionization, through observations of the HI 21\,cm forest against such a background source. 

\vspace{0.2cm}\noindent
{\it For more details see \citeauthor{smolcic15,afonso15,mcalpine15,makhathini15}  (\citeyear{smolcic15}, this Volume).}

\vspace{0.5cm}\noindent
{\bf $\diamond$ Diffuse Non-Thermal Emission from Galaxy Clusters and Cosmic Filaments} 

\vspace{0.2cm}\noindent
Radio Halos (RH) are known to be present in a high fraction of massive and merging clusters. The current prevalent view is that they trace turbulent regions, developed in merging clusters, where relativistic particles are trapped and accelerated. According to
such a scenario the synchrotron spectra of RH become increasingly steep above a certain frequency, whose value is determined by the efficiency of turbulent-acceleration.
Relaxed clusters are expected to host the so-called ``off-state'' RH, produced by secondary particles, that can be potentially used to detect galaxy clusters in blind radio surveys, providing a powerful cosmological tool, in synergy with upcoming X-ray and Sunyev-Zeldovitch (SZ) surveys (also possible with SKA1 at Band 5). Such off-state RH are predicted by modelling but not yet observed, due to lack of sensitivity. 

\vspace{0.2cm}\noindent
The phenomenon of mini radio halos, on the other hand, seems to be closely related to AGN feedback in galaxy clusters.  Mini-halos are detected around a number of radio-loud brightest cluster galaxies (BCG), on scales comparable to that of the cooling region. Mini-halos are not directly connected with radio bubbles, but the emission is on larger scales and is truly generated from the intra-cluster medium (ICM). 

\vspace{0.2cm}\noindent
Predictions at low and mid frequencies show that 
the SKA1 will step into an unexplored territory allowing us to study the formation and evolution of radio halos in a totally new range of cluster masses, dynamical status and redshift. 
Ongoing simulations of SKA1 observations of galaxy cluster halos show promising results, in particular thanks to the development of deconvolution and source detection algorithms optimized for the analysis of extended and diffuse radio sources.

\vspace{0.2cm}\noindent
Even more excitingly the SKA1 may, for the first time,  detect synchrotron emission produced by stationary accretion shocks in the warm-hot intergalactic medium associated with cosmic filaments. This is a very challenging task, but even a non-detection at SKA1 sensitivities will give an extremely important upper-limit on the magnetization of the Cosmic Web on its largest scales.

\vspace{0.2cm}\noindent
{\it For more details see \citeauthor{cassano15,ferrari15,gitti15,grainge15,vazza15} (\citeyear{cassano15}, this Volume).}

\vspace{0.5cm}\noindent
{\bf $\diamond$  Detailed Astrophysics of Star Formation and Accretion Processes} 

\vspace{0.2cm}\noindent
SKA1 continuum surveys covering a wide range of frequencies will provide $\mu$Jy sensitivities with exquisite image fidelity over a wide range of spatial scales for all nearby galaxies. This will produce a complete census of star-formation and AGN activity as a function of galaxy mass, morphology and spectral type, black-hole mass and luminosity.
This, alongside UV/optical/IR surveys (see \S6)
will be the cornerstone of multi-wavelength studies of the Local Universe. By providing high resolution ($0.5''$ or better at $\sim1\,$GHz) the SKA will be able to decompose individual
galaxies into their compact radio source populations (accretion dominated AGN, compact HII regions, superstar clusters, X-ray binaries, planetary nebulae, supernovae and their remnants). 

\vspace{0.2cm}\noindent
The high resolution provided by the SKA1, even over wide areas, will mean 
that high-resolution, multi-frequency total intensity and polarization imaging 
of a large sample of RL AGN will become available. Our
developing understanding of the dynamics and the spectral/polarization evolution of RL AGN means that it will be
possible to make relatively accurate estimates of the kinetic luminosity
function of the AGN and their environmental properties from radio observations alone. The ability to do this will be very important both for our understanding of the physics of the RL AGN population and for models of galaxy formation and evolution that rely on feedback processes from RL objects. 

\vspace{0.2cm}\noindent
The main physical processes at work in RQ AGN, particularly the origin of the 
radio emission, are still under debate, 
mainly due to the weakness of these objects. So far high-resolution studies have been possible only for local RQ AGN where high brightness temperature radio cores have been detected. The main processes proposed to explain such emission are synchrotron radiation from mildly relativistic mini-jets, thermal cyclo-synchrotron emission by low-efficient accretion flow, or thermal free-free emission from the X-ray heated corona or wind. Multi-frequency, high-sensitivity polarimetric radio observations are, thus, crucial to constrain the nature of the power engine, and may help in distinguishing between the contribution from star formation and AGN activity. 

\vspace{0.2cm}\noindent
{\it For more details see  \citeauthor{beswick15,kapinska15,orienti15,wolter15} (\citeyear{beswick15}, this Volume).}

\vspace{0.5cm}\noindent
{\bf $\diamond$ Strong Gravitational Lensing} 

\vspace{0.2cm}\noindent
Strong gravitational lenses (GL) provide an important tool to measure masses in the distant Universe, thus (a) testing models for galaxy formation and dark matter, (b) investigate structure at the Epoch of Reionization, and (c) measuring the Hubble constant, and possibly the $w$ parameter (from the cosmology equation of state), as a function of redshift. The limiting factor in all of these studies is the currently small samples of known GL ($\sim 10^2$).  SKA will transform our understanding of the Universe with gravitational lensing,  as the number of radio-detected GL will increase to $\sim 10^5$.  Depending  on splitted image separation, relative brightness and total flux density,  between $10^3$ and 
$10^4$ of such systems will be identified in an all-sky SKA1 mid-frequency survey. This requires a large area, sub-arcsecond angular resolution survey. SKA1 lensing simulations show that  angular resolutions of $0.25-0.5$ arcsec are required for identifying most of the lens candidates for follow-up imaging.  Resolution of $\le0.2''$ is required to follow-up these sources.

\vspace{0.2cm}\noindent
{\it For more details see \citeauthor{mckean15} (\citeyear{mckean15}, this Volume).}

\vspace{0.5cm}\noindent
{\bf $\diamond$ Legacy Value/Serendipity/Rare}

\vspace{0.2cm}\noindent
Experience shows that when telescopes enter unexplored areas of observational phase space, they make unexpected discoveries. Even more importantly, experience shows that the most significant discoveries with major telescopes are unexpected. For example, of the top ten discoveries with {\it HST}, only one was listed amongst the key goals used to justify {\it HST}. So while specific science goals are useful to focus the SKA design, they are unlikely to appear amongst its greatest scientific achievements. So in addition to achieving known science goals, the SKA must be designed to maximise its ability to discover the (potentially more important) unknown science goals. 
Moreover, the sensitivity of SKA1 means that all-sky surveys will result in a radio measurement of most galaxies detected in upcoming all-sky optical surveys. So SKA1 will transform radio-astronomy from a niche field focused on RL AGNs to a mainstream field producing data on most objects of interest by astronomers at other wavelengths. 

\vspace{0.2cm}\noindent
One particular area where the legacy added-value of an all-sky survey is significant is Galactic science. The SKA1 unique spatial resolution, sensitivity and survey speed at 1-2 GHz, will provide us with a valuable and unprecedented wide-field atlas of the Galactic continuum emission. SKA1 will give the opportunity to create a sensitive catalog of discrete Galactic radio sources, most of them representing the interaction of stars at various stages of their evolution with the environment: complete census of all stage of HII regions evolution; complete census of late stages of stellar evolution such as PNe and SNRs; detection of stellar winds, thermal jets, Symbiotic systems, Chemically Peculiar and dMe stars, active binary systems in both flaring and quiescent states. Coherent emission events as Cyclotron Maser due to particle acceleration and interaction with exoplanets can be detected. 

\vspace{0.2cm}\noindent
{\it For more details see \citeauthor{norris15,umana15} (\citeyear{norris15}, this Volume).}

\begin{table}
\footnotesize
\centering
  \caption[Landscape table]{Outline of {\it Reference} Surveys}  
  \label{tab:sur}
 \begin{tabular}{@{} ccclcccc @{}}
  \hline
\multicolumn{1}{l}{\multirow{2}{*}{\#}}  & \multicolumn{1}{l}{\multirow{2}{*}{\parbox{1cm}{Science Drivers}}} & $\nu_{\rm obs}$ &  Tier & rms & Area & $\theta$  &   \multicolumn{1}{l}{\multirow{2}{*}{\parbox{5cm}{Increment over pre-SKA }}}   \\ 
  & & (GHz)  &  &  ($\mu$Jy/b) & (deg$^2$) & ($''$)  & \\ 
   \hline
& & & & & & &  \\
\multicolumn{1}{c}{\multirow{5}{*}{1}} & \multicolumn{1}{c}{\multirow{6}{*}{\parbox{2cm}{SFHU SMBH/gal co-evolution}}} & \multicolumn{1}{c}{\multirow{6}{*}{$\sim 1^a$}} & \multicolumn{1}{c}{\parbox{1cm}{Ultra Deep}} & 0.05 & $1$ & $0.5^b$  &  
\multicolumn{1}{l}{\parbox{5cm}{$40\times$ deeper than VLASS-3; $10\times$ smaller area; similar resolution}}\\
 & & & & & & &  \\
  &        &  & \multicolumn{1}{l}{Deep} & 0.2 & 10-30 &  0.5   & \multicolumn{1}{l}{\parbox{4.5cm}{$10\times$ deeper than VLASS-3; same area; similar resolution}} \\ 
 & & & & & & &  \\
     &      &  & Wide & 1 & \multicolumn{1}{l}{\parbox{1cm}{1 $10^3$}} & 0.5  & \multicolumn{1}{l}{\parbox{5cm}{Same sensitivity as MIGHTEE-2 survey; $30\times$ larger area; $8\times$ higher resolution}}\\
&  & & & & & &  \\
\hline
 & & & & & & &  \\
\multicolumn{1}{c}{\multirow{5}{*}{2}} & \multicolumn{1}{c}{\multirow{5}{*}{\parbox{2cm}{SFHU SMBH/gal co-evolution}}}   & \multicolumn{1}{c}{\multirow{5}{*}{$\sim 10$}}  & \multicolumn{1}{c}{\parbox{1cm}{Ultra Deep}} & 0.04 & 0.0081 & $0.1$  &  \multicolumn{1}{l}{\parbox{5cm}{$20\times$ deeper than JVLA 8 GHz GOODS-N field; similar area; 2$\times$ better resolution}}\\
   &   &  & &   &  &   &   \\ 
&  &  & Deep & 0.3 & 1 &  0.1   &  \multicolumn{1}{l}{\parbox{5cm}{$2\times$ deeper than 5 GHz tier of  eMERGE legacy survey; $20\times$ larger area; similar resolution}}\\
 &     &  & &   &  &   &  \\  
\hline
  &    &  & &   &  &   &  \\  
3 & \multicolumn{1}{l}{\parbox{1.65cm}{Clusters and Filaments}}  & 0.12 & All-sky & $\sim 20^c$ & 31 $10^3$  & 10  &   \multicolumn{1}{l}{\parbox{5cm}{3$\times$ better surface brightness sensitivity than LOFAR all sky surveys, corresponding to the detection of 10$\times$ fainter radio halos/relics}}\\
&         &  &  &  &  &   &  \\
\hline
 & & & & & & &  \\
\multicolumn{1}{c}{\multirow{5}{*}{4}}  & \multicolumn{1}{l}{\parbox{2cm}{Strong GL}} & \multicolumn{1}{c}{\multirow{5}{*}{1.4}}  & \multicolumn{1}{c}{\multirow{5}{*}{All-sky}} & \multicolumn{1}{c}{\multirow{5}{*}{4}} & \multicolumn{1}{c}{\multirow{5}{*}{31 10$^3$}} & $\leq 0.5^d$  &  \multicolumn{1}{l}{\parbox{5cm}{$2-3\times$ deeper sensitivity than ASKAP all sky survey (EMU); $20\times$ better angular resolution; $\sim 1000\times$ more radio-loud strong GL than currently known}}\\
   &  &  &  &  & &   & \\ 
 &  \multicolumn{1}{l}{\parbox{2cm}{ Legacy/Rare Serendipity}} &  &  &  &  & $\sim 2^d$  & \multicolumn{1}{l}{\parbox{5cm}{$2-3\times$ deeper sensitivity than ASKAP all sky survey (EMU); $5\times$ better angular resolution }}\\
  &   &  &  &  & &   & \\ 
   \hline
\multicolumn{8}{l}{$^a$ Reference value. The observing frequency can be fine-tuned within  Band 1 and/or  2}\\
 \multicolumn{8}{l}{$^b$ Reference value at 1 GHz. $<1$ arcsec required to avoid confusion (see text)}\\
 \multicolumn{8}{l}{$^c$ may be confusion limited (see Appendix)}\\
 \multicolumn{8}{l}{$^d$ Different angular resolution requirements reflect in distinct surveys,  unless we can assume that the same survey can be}\\
 \multicolumn{8}{l}{\hspace{0.5cm}  processed at both the required resolutions} 
  \end{tabular}

\end{table}

\begin{figure}
  \centering
  \includegraphics[width=.52\textwidth]{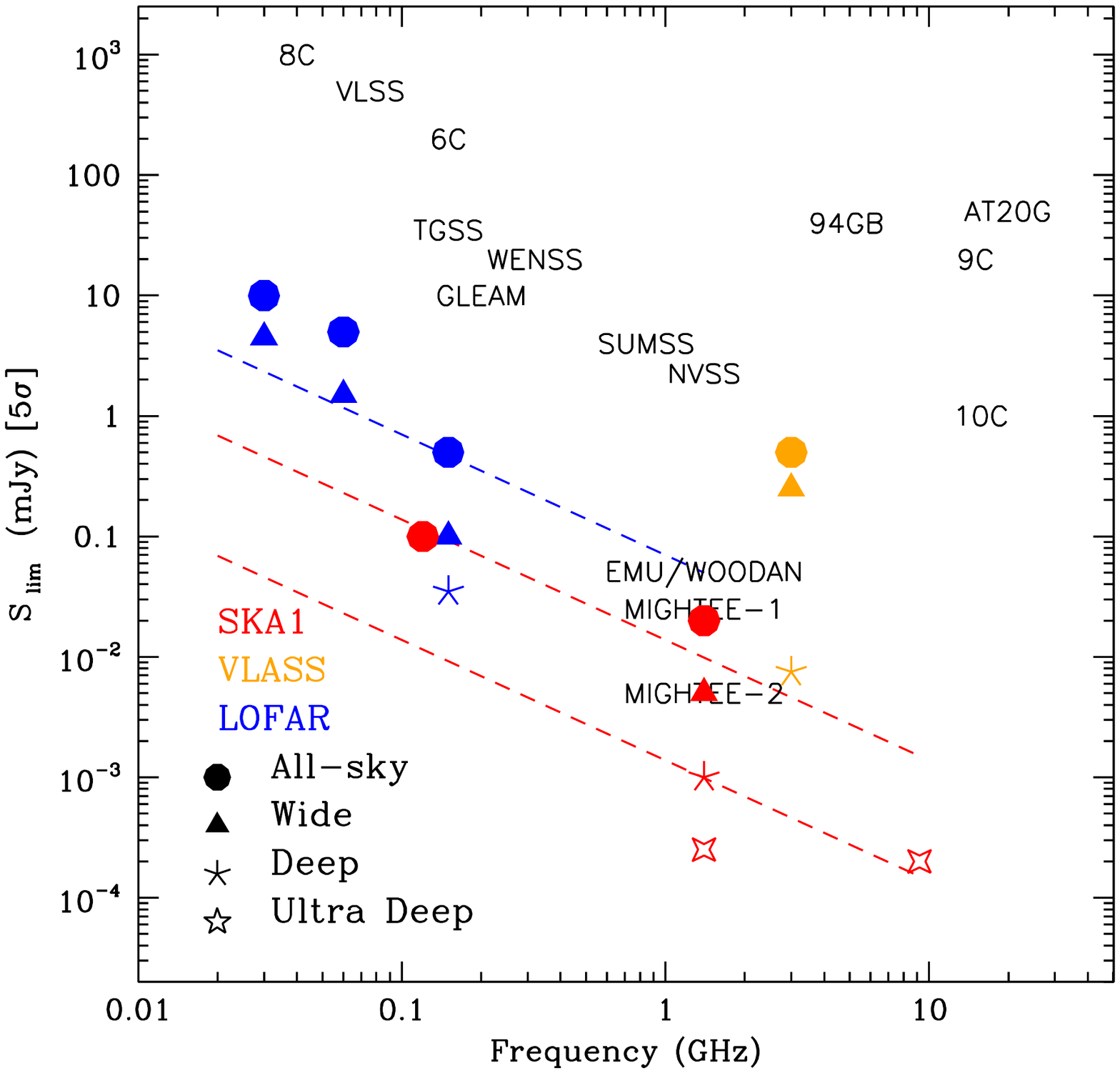}\includegraphics[width=.5\textwidth]{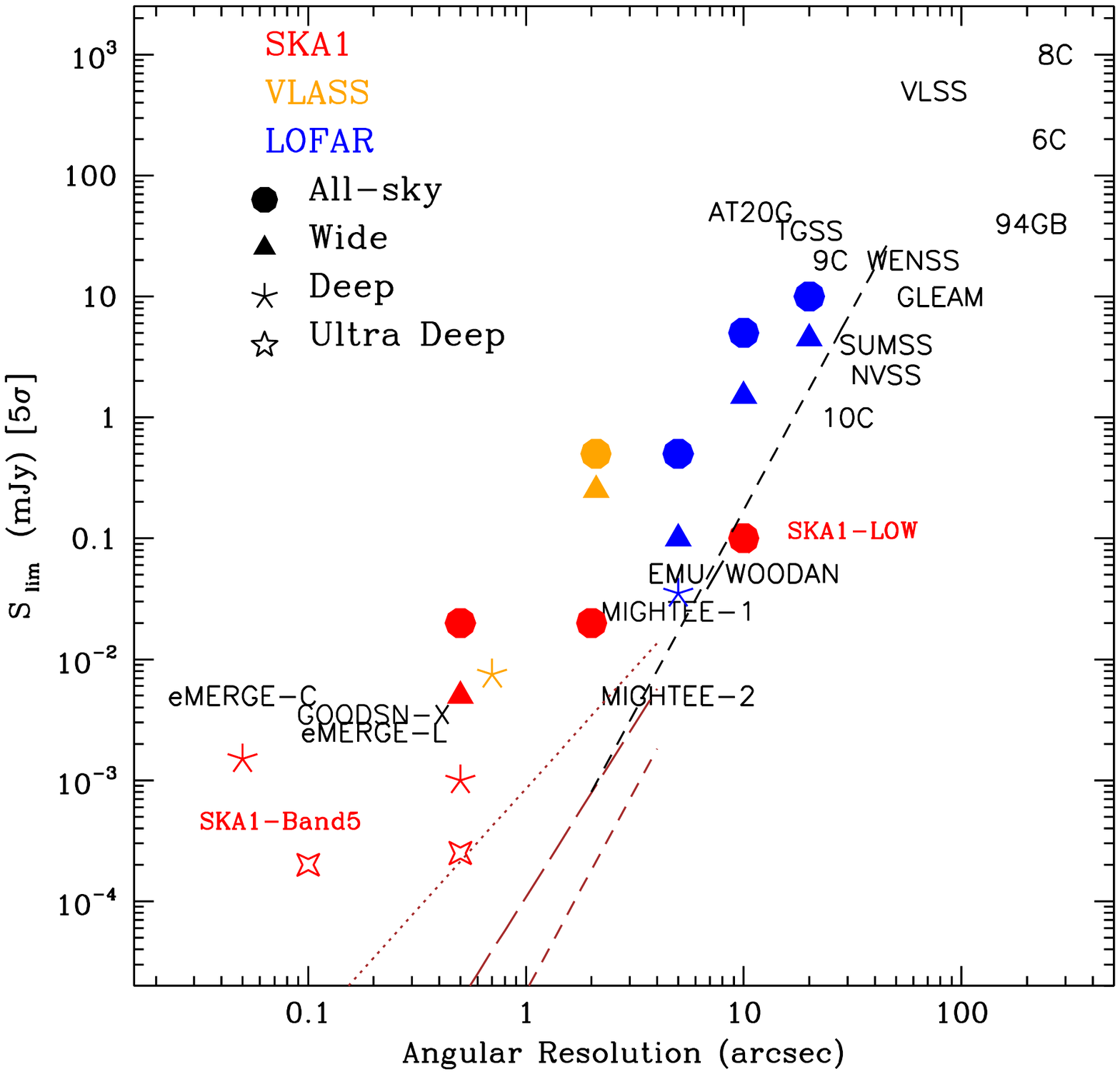}
  
  \vspace{-3cm}
  \includegraphics[width=.52\textwidth]{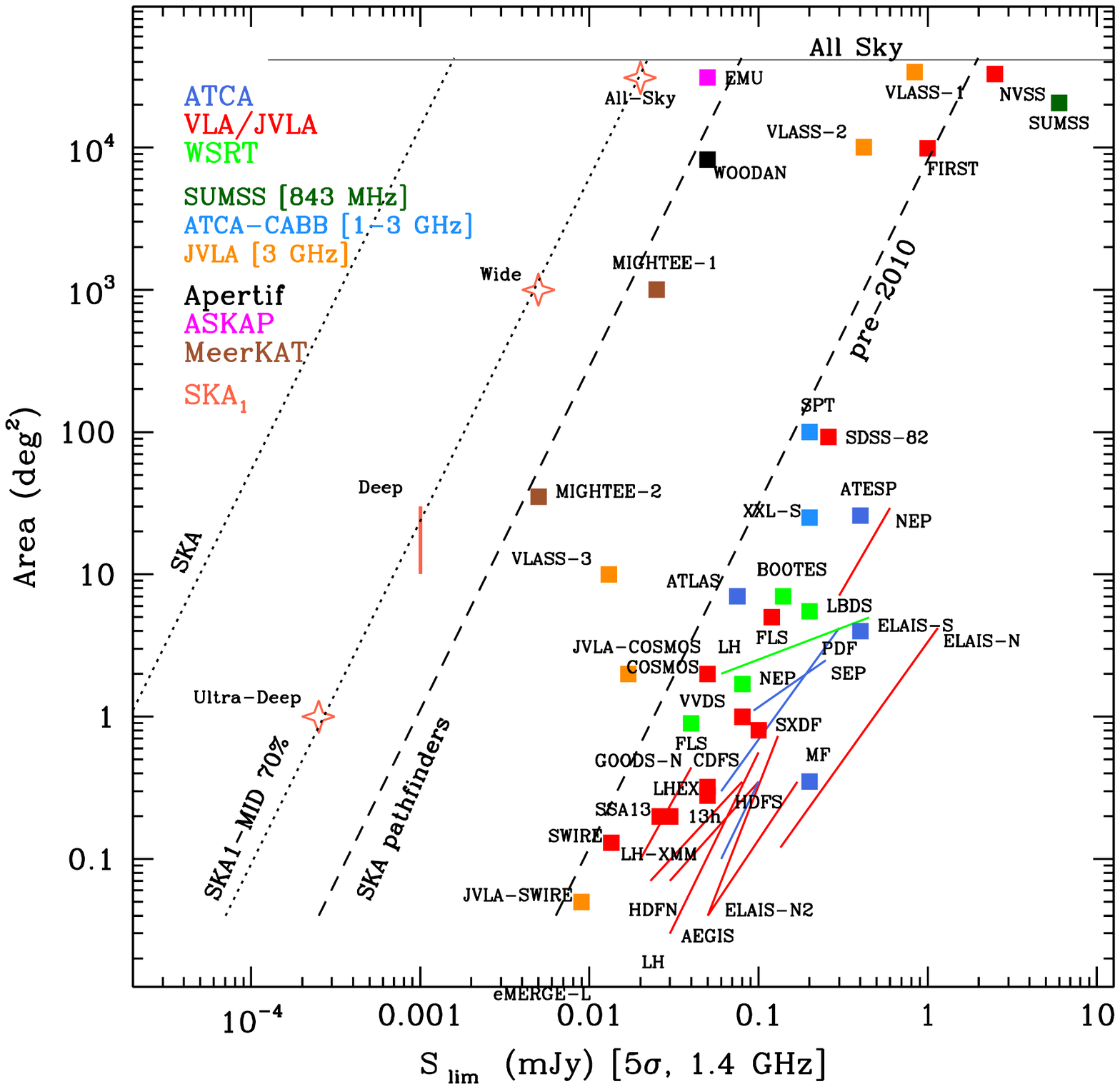}

  \vspace{-2cm}
\caption[]{{\it Top Panels:} SKA1 {\it reference} surveys in comparison with existing $>2\pi\,$steradian surveys and/or surveys planned for the next future with SKA pathfinders and precursors. LOFAR, VLASS and SKA1 {\it reference} surveys are highlighted in blue, orange and red respectively. Different symbols refer to different a survey coverage: all-sky (filled circles); wide tiers (filled triangles); deep tiers (asterisks); ultra deep tiers (starred symbols). {\it Left:} Depth (5$\sigma$ flux limit) vs. frequency. Band 1 and/or 2 SKA surveys are all shown at a reference frequency of 1.4 GHz. The red and blue dashed lines indicate a slope of $\sim \nu^{-1}$ for different 1.4 GHz flux normalizations. {\it Right:} Depth (5$\sigma$ flux limit) vs. angular resolution. The black and brown  lines represent approximate estimates of the confusion limit at 120 MHz and 1 GHz respectively (see Appendix for more details). {\it Bottom panel:} SKA1 {\it reference} surveys in comparison with existing or planned surveys. Only surveys with observing frequencies in the range 1-3 GHz are shown.  Area coverage vs depth (5$\sigma$ flux limit); for 3 GHz VLA surveys the flux limit has been rescaled to 1.4 GHz. }
  \label{fig:Survey}
\end{figure}

\section{Reference Surveys}


\noindent 
We have identified four {\it reference} surveys for SKA1, that we consider best suited to address the aforementioned top priority science cases, as listed below:
\begin{itemize}
\item Three-tiered survey at Band 1 or 2 
\item Two-tiered survey at Band 5
\item All-sky survey with SKA1-LOW
\item All-sky survey at Band 2 
\end{itemize}
\noindent
Their observational parameters (observing frequency, 1$\sigma$ rms sensitivity, surveyed  sky area, angular resolution) are summarized in Table~\ref{tab:sur}, as well as their main science drivers. Such surveys are defined so as to provide significant advances over pre-SKA surveys (as illustrated in the last column of Table~\ref{tab:sur}).
The {\it reference} surveys are compared to existing and  planned surveys (with SKA pathfinders and precursors) in Figure~\ref{fig:Survey}: the three panels illustrate the  increment over a pre-SKA scenario in terms of flux limit (top right), angular resolution (top left) and area covered (bottom). 

\noindent
Our reference surveys were originally based on full sensitivity capabilities as  advertised in 
\cite{Dewdney13} and \cite{braun14}\footnote{Such sensitivity performance is summarized in the Appendix.}, by assuming 30\% fractional bandwidths (i.e. $BW=0.3\nu$, where $\nu$ is the observing frequency). Since SKA1-SUR has been deferred as a result of the re-baselining process, all Band 1/2 surveys now refer to SKA1-MID.  In addition we now fully exploit the large bandwidths available on MID. This change allows us to fully compensate the factor $\sim2$ longer integration times due to the  30\% reduced collecting area of SKA1-MID introduced by the re-baselining, at least for the high resolution (sub-arcsec) surveys. The low resolution Band 2 all-sky survey, on the other hand, would be better done with SKA1-SUR (a factor $\sim 4$ faster at a few arcsec resolution).  This means that the original case for a 2 $\mu$Jy, 2 arcsec resolution survey with SKA1-SUR (extensively discussed in \citealt{norris15}, this Volume), is not achievable, and a 4 $\mu$Jy rms is now quoted.
We also note that the 50\% reduced collecting area of SKA1-LOW implies a factor 4 longer timescale for the (confusion-limited) all-sky survey, bringing the original 6 months integration time to 2 years.  
Similar timescales are also needed for the Band 1 and/or 2 surveys, while the Band 5 survey could be 
in principle run in 6 months. It is worth noting, however, that the actual survey parameters can be fine-tuned to a certain degree to meet tighter time constraints and dependent on refined science goals. For example, the use of the wide bandwidth introduces non-uniform sensitivity as a function of frequency due to the varying primary beam. 

\begin{table}
\footnotesize
\centering
  \caption[Landscape Table]{Tiered surveys at Band 1/2 and at Band 5}  \label{tab:sur2}
 \begin{tabular}{@{} l|l|l @{}}
   \hline
 Survey & \multicolumn{1}{l}{Science Requirements} & \multicolumn{1}{|l}{Addressed Science}  \\ 
\hline
 & & \\
1 - UD & \multicolumn{1}{l|}{\parbox{10cm}{will detect $SFR>10 M_{\odot} /yr$ galaxies up to $z\sim 3-4$, and $SFR>50 M_{\odot} /yr$ galaxies up to $z\sim 6$ to probe the SFHU in the very early phases of galaxy evolution }} 
&
\multicolumn{1}{l}{\multirow{6}{*}{\parbox{3cm}{W/D/UD: SFHU non-thermal \\ W/D/UD: SMBH/gal co-evolution \\ W: SF/AGN astrophysics and resolved SF in nearby Universe  \\ W/D: RL/RQ AGN dichotomy}}}\\
 & & \\
1 - D & \multicolumn{1}{l|}{\parbox{10cm}{will detect $SFR>10 M_{\odot} /yr$ galaxies at $z\sim 1-2$, the epoch of maximum star formation activity, for detailed SFG studies as a function of stellar mass, environment and galaxy morphology}} & \\
 & & \\
1 - W & \multicolumn{1}{l|}{\parbox{10cm}{will probe the bulk of the SFG population ($0.5<\rm{SFR}<10$ M$_{\odot}$/yr) over a wide range of environments, for detailed studies as a function of stellar mass, environment and galaxy morphology, in the redshift range where the evolution function presents a strong derivative ($0<z<1$)}} & \\
 & & \\
\hline
 & & \\
2 - UD & \multicolumn{1}{l|}{\parbox{10cm}{- will detect galaxies forming stars at a rate of  $\sim 50 M_{\odot}/yr $ out to $z\sim 6$ in their rest-frame 70 GHz emission;\\ 
- will map the distribution of star formation on kpc ($\lesssim 0.2$ arcsec) scales in galaxies forming stars at  $\sim 100 M_{\odot}/yr $ out to the peak of the cosmic star formation rate density ($z\lesssim 2$; rest-frame 30 GHz); \\
- will resolve  $SFR>100 M_{\odot}/yr $ galaxies on sub-kpc ($\sim 0.1$ arcsec) scales up to $z\sim 1$ to map out the distribution of active star formation and morphologically separate out AGN contributions to total energetics for unbiased measurements of SFRs, and detailed studies of AGN-galaxy co-evolution/feedback }} &
  \multicolumn{1}{l}{\multirow{4}{*}{\parbox{3cm}{D/UD: SFHU thermal and resolved SF at medium/high-z  \\  D/UD: SF/AGN Interplay}}}\\
 & & \\
2 - D & \multicolumn{1}{l|}{\parbox{10cm}{- will detect galaxies forming stars at a rate of $\sim 100 M_{\odot}/yr $ out to $z\lesssim 3$ (rest-frame 40 GHz) \\
- will resolve $SFR>100 M_{\odot} /yr$ galaxies on sub-kpc ($\lesssim 0.1$ arcsec) scales out to $z\sim 0.5$ to map out the distribution of active star formation and morphologically separate out AGN contributions to total energetics for unbiased measurements of SFRs, and detailed studies of AGN-galaxy co-evolution/feedback}} & \\
 & & \\
\hline

  \end{tabular}

\end{table}

\begin{table}
\footnotesize
\centering
  \caption[Landscape table]{All-sky survey with SKA1-LOW}  \label{tab:sur3}
 \begin{tabular}{@{} l|l @{}}
   \hline
 Science Requirements & Addressed Science  \\ 
  \hline
 &  \\
\multicolumn{1}{l|}{\parbox{11cm}{- will detect $\sim 2600$ RHs up to $z\sim 0.6$, including $\sim 1000$ USSRH, a factor $\sim 7\times$ increase over LOFAR All Sky Surveys\\ 
- will probe for the very first time the existence of hadronic 'off-state' radio halos, predicted for virialized clusters; \\
- will be able to detect RHs in clusters up to $z\sim 1$ and with masses down to $M_{500} \sim 10^{14} M_{\odot}$, thus competing with {\it eROSITA} and SZ surveys in blind cluster searches; \\
- will detect $>1-10\%$ fraction of the diffuse non-thermal emission in massive $B\sim 0.1-0.3$ $\mu$G cosmic filaments, or provide an upper limit in case of lower magnetic fields; \\
- will find powerful RL AGN in the Epoch of Reionisation}} &
\multicolumn{1}{l}{\parbox{3cm}{RH and USSRH\\ hadronic 'off-state' RH\\ cosmic filaments\\ RL AGN physics\\ RL AGN in the EoR\\ rare populations}}\\
 &  \\
\hline
  \end{tabular}

\end{table}

\begin{table}
\footnotesize
\centering
  \caption[Landscape table]{All-sky Survey at Band 2}  \label{tab:sur4}
 \begin{tabular}{@{}  l|l|l @{}}
   \hline
Survey & Science Requirements & Addressed Science  \\ 
\hline
&& \\
\multicolumn{1}{l|}{\parbox{1.5cm}{4 $\mu$Jy rms \\ $<0.5''$ res.}} & \multicolumn{1}{l|}{\parbox{9.5cm}{- will detect $\sim 10^4$ new strong lens systems ($\sim 0.25\,$deg$^{-2}$) for VLBI follow-up; \\ 
- will resolve 100 pc scales at distances of $D\sim 40\,$Mpc, for detailed studies of ISM in nearby galaxies }} &
\multicolumn{1}{l}{\parbox{3cm}{Strong GL \\ SF/AGN astrophysics in nearby galaxies \\ rare populations}}\\
&& \\
\hline
&& \\
\multicolumn{1}{l|}{\parbox{1.5cm}{4 $\mu$Jy rms \\ $\sim 2''$ res.}} & \multicolumn{1}{l|}{\parbox{9.5cm}{- will provide a radio counterpart for most galaxies detected in upcoming all-sky optical surveys; a resolution of $\sim 2$ arcsec is optimized for radio-optical identification experiments;  \\
- will provide a complete and unprecedented census of Galaxy Plane radio sources (stellar winds, PN, HII, active binary systems, flare stars etc.), increasing the number of detections by a factor 100$\times$; \\
- will provide high-$\nu$ information for cluster non-thermal emission, by detecting $\sim 500$ new halos (in both merging and relaxed clusters) and $\sim 300$ new mini-halos (in cool core clusters) up to $z\sim 0.6$;    \\
- will provide  $200\times$ more radio sources than currently known (from 2.5 million to 500 million), allowing to identify rare populations and probably discovering unknown classes of objects }} &
\multicolumn{1}{l}{\parbox{3cm}{Legacy \\ Galaxy Plane  \\Clusters/mini-halos \\ Serendipity \\ rare populations }}\\
&& \\
\hline
  \end{tabular}

\end{table}

\subsection{Science Requirements and Critical Parameters}

\noindent 
The observational parameters  of our {\it reference} surveys have been chosen so as to respond to specific science requirements, as detailed In Tables~\ref{tab:sur2}, \ref{tab:sur3}, \ref{tab:sur4}, and as such they guarantee a significant impact in the research areas that we have identified as  top-priority. It s clear however that a certain degree of freedom is allowed to some of them. We believe it is very important to best fine-tune the observational parameters of SKA1 surveys, to fully exploit existing commensalities and synergies among different science areas (HI galaxy science, magnetism, cosmology, transients, etc.). The most critical observational parameters for our science objectives are discussed below. 

\paragraph{Observing Frequency: }
For most continuum science cases we are frequency agnostic as long as the requisite source density, galaxy star-formation rates, AGN radio powers are met.  Therefore, our quoted sensitivities (see Table~\ref{tab:sur}) should be scaled by the average spectral index of the dominant extragalactic source population at these flux densities, i.e. star-forming galaxies, which have a spectral index of $\nu^{-0.7}$. Hence, as long as the required resolution is retained, continuum surveys are generally better done at as low frequency as feasible, albeit before becoming dominated by the sky temperature. For the mid-frequency  arrays however, this should not be an issue and observing down to a frequency of$\sim  600-700$ MHz could be advantageous. It is worth  noting that for observations with a lower central frequency field-of-view, and hence survey speed, increases. Observing at low frequency would also exploit commensality with cosmological HI surveys, and provide HI line information (HI masses/redshifts) for at least some classes of galaxies. Observations at higher frequency (e.g. $\sim 1.4$ GHz), on the other hands, would exploit commensality with polarization surveys (magnetism). As extensively discussed by \citet[][this Volume]{murphy15},  we gain significantly by having Band 5 as second band, as this would provide complementary significantly higher resolution information ($0.05-0.1$ arcsec), very important for detailed astrophysical studies at high redshift and AGN identification (see below). In addition Band 5 allows  us to probe a different emission process at high redshift (thermal vs non-thermal). Band 5 is also crucial for Galactic studies, not only because of the high resolution and the spectral index information it provides, but also because it allows to detect particular classes of sources, like Hypercompact HII regions or a quiescent Solar analogue at 10\,pc (see \citealp{umana15}, this Volume). 

\paragraph{Angular Resolution at Mid/High Frequency: \\}
{\it $\bullet$ Galaxy Evolution: }
Angular resolution is a critical parameter for galaxy evolution studies for several reasons. The most obvious one is confusion from background sources. Our {\it reference} Ultra Deep and Deep Band 1/2 tiers would be confusion-limited at $\sim 1-2$ arcsec resolution, assuming the latest 1.4 GHz source counts predictions (\citealp{Condon12}; see Appendix for more details). Another critical aspect is the ability to identify and reliably classify different classes of sources at any redshift. This requires extensive, deep multi-wavelength information in the surveyed areas, as well as sub-arcsec resolution radio observations. Galaxy H$_{\alpha}$ disks are found to shrink to $0.7$ arcsec sizes at redshifts $z\gtrsim 1$ (based on HST Grism imaging; e.g. \citealp{Nelson13}), corresponding to $5-6$ kpc physical scales. So radio observations of similar or better resolution are needed to distinguish emission triggered by star formation from emission from $\gtrsim 10-30$ kpc radio jets possibly associated to RQ AGNs. Assuming a 0.5 arcsec resolution at 1 GHz (achievable with $\geq 120$ km maximum baselines), we allow for resolution values  from 0.4 (1.4 GHz)  to 0.7 (700 MHz) arcsec, depending on the actual observing frequency. Such values provide a good match with the resolution of next generation optical/NIR surveys (0.2 arcsec {\it Euclid}; 0.7 arcsec LSST). A factor $10-100$ better resolution would be needed to resolve star formation from embedded AGN radio cores. Galaxy-AGN co-evolution studies will significantly benefit from the implementation of Band 5 in phase 1 (0.05-0.1 arcsec at 10 GHz), that would allow to pinpoint flat-spectrum radio core candidates (see \citealp{mcalpine15}, this Volume). Crucial in this respect is  the implementation of VLBI capabilities for SKA1. 

{\it $\bullet$ Strong Lensing: }
The main requirement for gravitational lens surveys is resolution. Retaining the post-rebaselining 150 km maximum baseline for SKA1-MID is therefore crucial. For maximum efficiency, resolutions of about 0.2 arcsec are required, as was used in the CLASS survey. However, surveys of 0.5 arcsec resolution are still useful, at the expense of completeness of samples (the smaller-separation lenses will be missed) and false positives (requiring more follow-up). Nevertheless, this problem has been investigated in the past and tricks are available to find lenses with the lower-resolution surveys. A particularly promising approach is to use two surveys - here, SKA1 together with {\it Euclid} - as this should allow much more efficient rejection of false positives. In principle, 1 in every $\sim 600-1000$ background sources is strongly lensed, the main dependency being on source redshift (approximately as $(1+z)^4$); the likely redshift distribution of SKA objects suggests that this will be continued, although detailed simulations including the various source populations have yet to be done. A survey of a few $\times 10^8$ objects, e.g. all-sky with SKA1, should in theory yield a few hundred thousand lenses (\citealp{mckean15}, this Volume). SKA1 VLBI capability is critical to follow up the most promising strong lens candidates. 

\paragraph{Angular Resolution for SKA1-LOW:}
The lack of baselines longer than $\sim 80$ km for SKA1-LOW limits the overall sensitivity of the instrument which is confusion limited at $\sim 20$ $\mu$Jy/b rms (at 120 MHz, 10 arcsec resolution). Nevertheless SKA1-LOW (even with its post-rebaselining 50\% sensitivity performance) will provide a significant improvement in studies of cluster diffuse emission (\citealp{cassano15}, this Volume), at least up to $z\sim 0.5-1$ (at higher redshifts a 4-5 arcsec resolution become critical to reliably separating diffuse emission from contaminating point sources), and can potentially provide a first detection of the diffuse synchrotron emission in massive cosmic filaments (see simulations by \citealp{vazza15}, this Volume). For other key continuum science cases, better sensitivity and/or better resolution than currently designed, are necessary to make a significant step forwards with respect to SKA pathfinders (and in particular LOFAR, that already offers better spatial resolution). We therefore envisage the implementation of significanly longer baselines (ideally by a factor 4) for the full SKA. 

\paragraph{Beam Shape:}
At $\mu$Jy flux density levels the source counts are such that effective beam solid angle is extremely sensitive to extended (large $\Omega$) low-level ``pedestals'' and other sidelobes of the dirty beam. The naturally weighted beam of SKA1-MID, in particular, has large pedestals produced by the 1 km core array optimized for pulsars and the 8 km halo array for HI observations. Consequently it is important that the (u,v) be weighted sufficiently to remove the pedestals from the compact cores of the SKA arrays. It is also important to investigate further the effect of sidelobes to confusion noise which cannot be curtailed by weighting.

\paragraph{\bf Dynamic Range: }
The sensitivities possible over wide areas with SKA1 naturally raise the requirement of high dynamic range for even the smallest survey tier (this is closely related to the beam shape and sidelobes discussed above). While some effort can be made to avoid the brightest sources, that will not always be possible. The use of an accurate sky model can help mitigate this issue, but there will undoubtedly be regions of enhanced noise (above thermal) around bright source. For these to be minimized we require a dynamic range of $\sim 60$ dB for the survey scenarios outlined above, but ultimately $\sim 70$ dB will be required for the full SKA.    

\paragraph{Short Baselines: }
The detection of diffuse non-thermal emission in cosmic filaments needs the availability of good UV coverage at the shortest baselines, ie those sensitive to diffuse emission on scales $\gg 1$ Mpc (corresponding to angular scales of degrees at redshift $z\sim 0.05-0.1$). This requirement is fulfilled  for SKA1-LOW (and not for SKA1-MID or SUR), which therefore plays a unique role for this experiment. On the oher hand, current simulations show that better surface brightness sensitivity is needed to fully probe diffuse radio-continuum emission from the cosmic web (\citealp{vazza15}, this Volume). This will be achieved by the full SKA, with longer baselines. 

\paragraph{Spectral Index Information: }
Several continuum science cases benefit from spectral index measurements. This is essential to identify classes of very steep spectrum sources (non thermal emission in galaxy clusters, very high redshift AGNs, etc.). The proposed {\it reference} surveys at different frequencies have matched sensitivity for $\nu^{-1}$ sources (see Figure~\ref{fig:Survey}, top right panel).

\paragraph{Strategy: }
Here we note a few points on general survey strategy. The use of nested ``wedding-cake'' surveys reduces the overall observing time by 
not requiring the duplication of observation over the same areas. Hence a natural strategy is to incrementally build up depth in all-sky surveys,
but go deep first in the best understood areas where you plan to go deepest. 
Related to this approach is the fact that the deepest pointings will
be constrained by both the location of the best ancillary data as well as bright, dynamic-range limiting sources. As confusion noise is 
the ultimate constraint on any survey we advocate that long baselines be kept (over sensitivity) and built early to maintain the SKA's position as a transformational observatory. 

\paragraph{Data Archiving and Processing: }
The possibility to store visibilities and re-process the data at different resolutions/bandwidths etc. would have a huge commensality/synergy added-value for continuum surveys. At the very least, if visibilities are not retained, we must plan for multiple processing paths to produce images with different resolutions. 



\begin{table}
 \centering
  \caption{Table of key multi-wavelength surveys for SKA surveys in the 2020 era. 
   All magnitudes are given in AB. 
The final column indicates the estimated release 
   date of the data after survey completion.}
   \label{tab:anc}
   \begin{tabular}{@{} cccccc @{}}
\hline
\multicolumn{1}{c}{SKA}  & \multicolumn{1}{c}{Observatory}  & \multicolumn{1}{c}{Area}    & \multicolumn{1}{c}{Wavelength} & \multicolumn{1}{c}{Limiting}& \multicolumn{1}{c}{Data} \\
\multicolumn{1}{c}{Tier} & \multicolumn{1}{c}{/Survey}    & \multicolumn{1}{c}{($\deg^2$)} & \multicolumn{1}{c}{Bands}   & \multicolumn{1}{c}{Mag.} & \multicolumn{1}{c}{Release} \\
&\multicolumn{1}{c}{Name}&&&\multicolumn{1}{c}{or flux$^{\rm a}$}&\multicolumn{1}{c}{Date} \\
    \hline
All-sky & ALL{\it WISE}$^{\rm a}$ & 42,195 & 3.4, 4.6, 12, 22 $\mu$m & $70\,\mu$Jy & 2013\\
All-sky & PanSTARRS$^{\rm b}$ & 31k & $g$, $r$, $i$, $z$, $y$ & $r<24.0$ & 2020$^{\rm b}$\\
All-sky & LSST$^{\rm c}$ & 20k &$u$, $g$, $r$, $i$, $z$, $y$ & $r<27.5$ & 2020\\
All-sky & VISTA-VHS$^{\rm d}$ & 20 & $Y$, $J$, $H$, $K$ & $K<20.0$ &2011+\\
All-sky & {\it eROSITA}$^{\rm e}$ & $\sim42$k & $0.5-10\,$keV & $\sim10^{-14}\,$ergcm$^{-2}$s$^{-1}$ & 2018\\
All-sky & {\it EUCLID}$^{\rm f}$ & 15k & $0.55-2.0\,\mu$m+spec. & $YJH<24$ & 2020+\\
All-sky & TAIPAN$^{\rm g}$ & 20k & $0.37-0.87\,\mu$m spec. & $R<17.5$ & 2015\\
All-sky & 4MOST$^{\rm h}$ & 15-20k & $0.39-1.05\,\mu$m spec. & $r<22$ & 2019+\\
All-sky & MOONS$^{\rm i}$ & 15k & $0.8-1.8\,\mu$m spec. & (TBD) & 2019+\\
Wide & H-ATLAS$^{\rm j}$ & 570 & $70-500\,\mu$m & $S_{\rm 250\mu m}>44.5\,$mJy& 2015 \\
Wide & DES$^{\rm k}$ & 5000 & $g$, $r$, $i$, $z$, $y$ & $r<25$ & 2017\\
Wide & VISTA-Viking$^{\rm d}$ & 1500 & $Y$, $J$, $H$, $K_s$ & $K_s<21.2$ & 2012\\
Wide & VST-ATLAS$^{\rm d}$ & 4500 & $u'$, $g'$, $r'$, $i'$, $z'$ & $r'<22.2$& 2016\\
Wide & VST-KIDS$^{\rm d}$ & 1500 & $u'$, $g'$, $r'$, $i'$ & $r'<24.2$ & 2016\\
Wide & PanSTARRS Deep$^{\rm b}$ & 1200 & $0.5-0.8$, $g$, $r$, $i$, $z$, $y$ & $g<27.0$ & 2020\\
Deep & SCUBA2$^{\rm l}$ & $1/6$ & $450$/$850\,\mu$m & S$_{850\mu m}>3.5\,$mJy & ???? \\
Deep & HerMES$^{\rm m}$ & $1...270$ & $70-500\,\mu$m & $S_{\rm 250\mu m}>3.8...64\,$mJy & 2016\\
Deep & SERVS$^{\rm n}$ & 18 & $3.4,4.6\,\mu$m & $\sim2\,\mu$Jy & 2013\\
Deep & VISTA-VIDEO$^{\rm d}$ & 12 & $Z$, $J$, $H$, $K_s$ & $K_s<23.5$ & 2016\\
Deep & LSST (deep drilling)$^{\rm o}$ & 38.4 & (some of) $u$, $g$, $r$, $i$, $z$, $y$ & $r<30$ & 2020\\
Deep & UltraVISTA$^{\rm d}$ & 0.73 & $Y$, $J$, $H$, $K_s$, NB & $K_s<25.6$ & 2016\\
Deep & DES$^{\rm k}$ &  & & & \\
     \hline
\multicolumn{6}{l}{\small $^{\rm a}${http://wise2.ipac.caltech.edu/docs/release/allwise/}}\\
\multicolumn{6}{l}{\small $^{\rm b}${http://www.ps1sc.org}}\\
\multicolumn{6}{l}{\small $^{\rm c}${http://www.lsst.org/files/docs/LSSToverview.pdf }}\\
\multicolumn{6}{l}{\small $^{\rm d}${http://www.eso.org/sci/observing/PublicSurveys/sciencePublicSurveys.html} }\\
\multicolumn{6}{l}{\small $^{\rm e}${http://www.mpe.mpg.de/erosita}}\\
\multicolumn{6}{l}{\small $^{\rm f}${http://www.euclid-ec.org/}}\\
\multicolumn{6}{l}{\small $^{\rm g}${http://www.taipan-survey.org/}}\\
\multicolumn{6}{l}{\small $^{\rm h}${http://workshop.4most.eu/doc/4most\_poster\_121019s.pdf} }\\
\multicolumn{6}{l}{\small $^{\rm i}${http://www.roe.ac.uk/~ciras/MOONS/VLT-MOONS.html} }\\
\multicolumn{6}{l}{\small $^{\rm j}${http://www.h-atlas.org/} }\\
\multicolumn{6}{l}{\small $^{\rm k}${http://www.darkenergysurvey.org/}}\\
\multicolumn{6}{l}{\small $^{\rm l}${http://www.jach.hawaii.edu/JCMT/surveys/Cosmology.html} }\\
\multicolumn{6}{l}{\small $^{\rm m}${http://hermes.sussex.ac.uk/} }\\
\multicolumn{6}{l}{\small $^{\rm n}${http://www.cv.nrao.edu/~mlacy/servs.html} }\\
\multicolumn{6}{l}{\small $^{\rm o}${http://www.lsst.org/News/enews/deep-drilling-201202.html} }\\
\end{tabular}
\end{table}

\section{Synergies and Commensalities}
\noindent
For many, if not most, of the science goals listed in \S2 multi-wavelength, 
ancillary data is key for a plethora of reasons, e.g. determining redshifts, 
measuring other properties. In Table~\ref{tab:anc} we list those
facilities and surveys which will best complement the proposed survey tiers
at the time the SKA1 surveys will be conducted. 
These include imaging and spectroscopic surveys from optical to far-IR 
wavelengths. Of note, but not included in Table~\ref{tab:anc}, are ALMA and
{\it JWST}, as detailed deep survey plans have not yet been defined for those
facilities. \\
A full exploitation of the existing synergies between the SKA surveys and those at other wavelengths is key  to maximise the science return from the continuum surveys and to generate a strong involvement in all astronomical communities. This requires 
more detailed examination of these other surveys and coordination with them
over issues such as depth and areas covered, although observations at different 
wavelengths have different observational constraints. By 
ensuring careful coordination of the SKA tiers with all these surveys 
we can provide the richest possible legacy value for the community. \\
Potential commensalities exist between the survey tiers proposed here and those 
proposed by other WGs, namely Cosmology, HI, Magnetism and Transients. While it is beyond
the scope of this document to examine this issue in detail we note that  
finding  a common ground between these different surveys will greatly decrease 
the total demand on SKA1 time for large projects as well as providing valuable 
ancillary radio data products. For example, the same set of
observations with the Australian SKA Pathfinder will provide different data 
products for the HI, continuum and polarisation legacy projects ensuring 
close collaboration of the different science teams. Hence, due 
to the huge benefit of commensality, we propose it be discussed in subsequent 
fora.

\section{Conclusions}

\noindent
We have highlighted the major science areas in which continuum surveys with 
SKA1, and ultimately the full SKA, will have. We have come up with a list of 
{\it reference} surveys which are designed to examine how we can address a wide 
range of science goals. Such {\it reference} surveys are just the starting point
for planning real surveys with SKA1 and considerable more conversation within
the community is required. However, they bring into sharp focus the technical
requirements and critical parameters which must be achievable, as well as the
need to consult other WGs in designing surveys which address more than 
just continuum science and make the most efficient use the world's most 
powerful radio telescope. \\
We finish by noting that surveys, while driven by headline science goals, can 
address an extremely wide range of science, and hence, be of benefit to a wide
section of the astronomical community. By presenting a transformational step
in sensitivities and capabilities well designed surveys with SKA1 will provide
huge science return for the widest possible number of astronomers.

\acknowledgments{\noindent NS is the recipient of an ARC Future
Fellowship.}

\section*{Appendix: SKA1 Sensitivity Performance }

\noindent
As a common reference we summarize here the  set of telescope parameters used to estimate the sensitivity of SKA1, as advertised in \citet{Dewdney13} and/or \citet{braun14} (see Table~\ref{tab:performance}. Using the parameters in Table~\ref{tab:performance} one can compute the thermal RMS sensitivity ($\Delta S$) of each array/band as follows:

\begin{equation}
\Delta S^{\rm Jy} = \frac{SEFD^{\rm Jy}}{\eta_S \sqrt{2\Delta \nu_{\rm Hz} \, t_{\rm s}}} = 1.31 \times 10^{-5} \frac{SEFD^{\rm Jy}}{\sqrt{\Delta \nu_{\rm MHz} \, t_{\rm h}}}\label{eq:rms}
\end{equation} 
where $\Delta \nu$ is the assumed instantaneous bandwidth ($\leq \Delta \nu_{\rm max}$), $t$ is the integration time in hours, and  the system efficiency $\eta_S$ is assumed equal to 0.9. It is important to note that the tabulated sensitivities are just reference values. Simulations (see e.g. Figure 2 of \citealp{braun14}) show that the actual sensitivities measured in CLEANed images are typically a factor of 1.5-2 larger for a range of restoring beams around natural resolution, and degrade very rapidly at the highest and lowest resolutions. This has to be taken into account when estimating the actual expected sensitivities.    \\

\begin{table}[tbp]
 \centering
  \caption{Telescope Performance assumed for SKA1 before rebaselining.}
  \label{tab:performance}
  \begin{tabular}{@{} cccccccrrr @{}}
    \hline
\multicolumn{1}{c}{Telescope} & \multicolumn{1}{c}{Band} &  \multicolumn{1}{c}{Freq. Range} & \multicolumn{1}{c}{$\Delta \nu_{\rm max}$$^a$} & \multicolumn{1}{c}{$\nu_0$} & \multicolumn{1}{c}{FoV$_{\rm eff}^b$} & \multicolumn{2}{c}{SEFD$^\ddagger$} & \multicolumn{2}{c}{$\Delta S^c$$^\ddagger$}\\ 
     &   &  GHz & MHz & GHz & deg$^2$ & \multicolumn{2}{c}{Jy}  & \multicolumn{2}{c}{$\mu$Jy h$^{-1/2}$}\\ 
    \hline
  SKA1-MID & 1a  & 0.35-1.05 & 700 & 0.7 &  1.4 & 3.5  & & 4.6 &\\ 
                   & 1b  & 0.58-1.02 & 440 & 0.7 &  1.4 & 2.8$^d$ &  & 3.7 &\\ 
                   & 2  & 0.95-1.76  & 810 & 1.4 &  0.35 & 1.7$^d$ &  & 2.2 & \\ 
                   & 3  & 1.65-3.05   & 1400 & 2.35 &  0.124 & 2.1 & & 2.8 & \\ 
                   & 4  & 2.8-5.18   & 2380 & 4.0 & 0.043  & 2.3  & & 3.0 & \\ 
                   & 5  & 4.6-13.8   & $2\times 2500$ & 9.2 & 0.0081  & 2.8 &  &  3.7 & \\ 
  SKA1-SUR & 1  & 0.35-0.9  & 500 & 0.625 &  61 & 11.8$^e$ & (16.3) &  15.5 & (21.4) \\ 
                   & 2  & 0.65-1.67  & 500 & 1.4 & 18 & 7.1$^e$ & (9.8) & 9.3 & (12.8)\\ 
                   & 3  & 1.5-4.0   & 500 & 2.8 &  3.1 & 9.4$^e$ & (13.0) &  12.3 & (17.0)\\ 
  SKA1-LOW &   & 0.05-.35  & 250 & 0.05 &  39 & 19  & & 24.9$^f$ & \\ 
                   &   &   &  & 0.11 &  8 & 2.9  & & 3.8$^f$ & \\ 
                   &   &    &  & 0.16 &  3.8  & 2.6 & & 3.4$^f$ & \\ 
                   &   &    &  & 0.22 &  2.0  & 2.6 & & 3.4$^f$ & \\ 
    \hline
\multicolumn{10}{l}{$^\ddagger$ require an increase of 30\% for MID and 50\% for LOW post-rebaselining. }\\
\multicolumn{10}{l}{$^a$ Maximum instantaneous bandwidth. }\\
\multicolumn{10}{l}{$^b$ {\it Effective} FoV, i.e. the FoV you have to sum up to cover areas larger than the primary beam with }\\
\multicolumn{10}{l}{$\;\;$ uniform noise. For MID and LOW FOV$_{\rm eff}=0.5 \, \Theta_P$, where $\Theta_P$ is the Gaussian primary beam}\\
\multicolumn{10}{l}{$\;\;$  in square~degrees ($\Theta_P=(\pi/4)(66\lambda_0/d_{\rm dish/station})^2$). For LOW we assume $d_{\rm station}=45$ m.  For }\\
\multicolumn{10}{l}{$\;\;$ SUR FOV$_{\rm eff}=$FOV$_{\rm PAF}= 36 \, \Theta_P$, where $\Theta$ is computed as above, assuming $\lambda_0=\lambda_{\rm min}$, i.e. the  }\\
\multicolumn{10}{l}{$\;\;$ minimum wavelength in the band.}\\
\multicolumn{10}{l}{$^c$ RMS sensitivity reached in 1$^h$ of integration assuming the tabulated SEFD and $\Delta \nu=$100 MHz}\\
\multicolumn{10}{l}{$\;\;$ (see eq.~\ref{eq:rms}). }\\
\multicolumn{10}{l}{$^d$ Includes Meerkat dishes.}\\
\multicolumn{10}{l}{$^e$ Includes ASKAP dishes. Note that ASKAP can be included in only one of the three available }\\
\multicolumn{10}{l}{$\;\;$ bands, at your choice. For the other bands  you have to use SEFD values for SUR dishes only, }\\
\multicolumn{10}{l}{$\;\;$ (reported in parenthesis in the table).} \\
\multicolumn{10}{l}{$^f$ Note that with the current Baseline Design (maximum baseline  $\sim 90$ km), SKA1-LOW is }\\
\multicolumn{10}{l}{$\;\;$ confusion limited and cannot reach the tabulated noise levels (see text for more details.} \\
  \end{tabular}

\end{table}

Please note that as a consequence of the re-baselining process SKA1-MID and SKA1-LOW have collecting areas reduced by a factor 30 and 50 per cent respectively. This means that the $\Delta S$ quoted for SKA1-MID and SKA1-LOW in Table~\ref{tab:performance} should be increased by the same factor.  We also notice that SKA1-SUR has been deferred, and is not relevant anymore for the definition of our continuum surveys.  
Another important parameter to take into proper account when designing a deep continuum survey, and strictly related to spatial resolution is ({\it classical}) confusion noise, which limits the actual depth of a survey, due to the presence of unresolved background sources, below the detection limit. The confusion limit critically depends on the assumed sources counts (typically approximated by a power-law, $n(S)\sim k S^{-\gamma}$). 
We used the scaling relation presented in \citet[][see their eq.~27]{Condon12} to estimate 120 MHz and 1 GHz confusion limits  as a function of angular resolution (see black and brown short-dashed lines in Figure~\ref{fig:Survey}, top right panel). This scaling relation assumes $\gamma=1.6$. For 1 GHz surveys we also show other two estimates: one (brown long-dashed line) obtained by converting to 1 GHz the $\gamma=1.7$ power-law fit to the 3 GHz source counts recently derived for the JVLA SWIRE field \citep{Condon12}; the other one obtained by using an older scaling relation presented in \cite{Condon87}, where $\gamma=2$ (brown dotted  line). The latter estimate, which better match confusion limits for existing low frequency surveys (see \citealp{cassano15}, this Volume), is shown to illustrate a worst-case scenario. It is clear that SKA1-LOW surveys are likely to be confusion limited at the angular resolutions accessible with $\sim 90$ km maximum baselines. At Band 1/2 confusion is less of a problem, and deep surveys at 1-2 arcsec resolution should be feasible, even though there is a clear intrinsic uncertainty in the extrapolation of the measured number counts to sub-$\mu$Jy sensitivities. For a more comprehensive discussion of confusion we refer to \citet[][this Volume]{zwart15}, where different estimates are presented (see their \S~6).


\begin{thebibliography}{99}

\bibitem[Afonso et al.(2015)]{afonso15}{Afonso J., et al.\ 2015, "Identifying the first generation of radio powerful AGN in the Universe with the SKA", in proceedings of  "Advancing Astrophysics with the Square Kilometre Array",  \pos{PoS(AASKA14)071}}

\bibitem[Beswick et al.(2015)]{beswick15}{Beswick R.~J., et al.\ 2015, "SKA studies of nearby galaxies: star-formation, accretion processes and molecular gas across all environments" , in proceedings of  "Advancing Astrophysics with the Square Kilometre Array", \pos{PoS(AASKA14)070}}

\bibitem[Braun(2014)]{braun14}{Braun R.\ 2014, "SKA1  Imaging Science Performance Document", SKA Document, Rev A Draft 5}

\bibitem[Cassano et al.(2015)]{cassano15}{Cassano R., et al.\ 2015, "Cluster Radio Halos at the crossroads between astrophysics and cosmology in the SKA era", in proceedings of  "Advancing Astrophysics with the Square Kilometre Array",  \pos{PoS(AASKA14)073}} 

\bibitem[Ciliegi \& Bardelli(2015)]{ciliegi15}{Ciliegi P. \& Bardelli S.\ 2015, "Synergistic science with Euclid and SKA: the nature and history of Star Formation", in proceedings of  "Advancing Astrophysics with the Square Kilometre Array",  \pos{PoS(AASKA14)150}} 

\bibitem[Condon(1987)]{Condon87}{Condon J.~J.\ 1987, in Proceedings of the Arecibo UpgradingWorkshop, ed. J. H. Taylor \& M. M. Davis (Arecibo:NAIC), 89}

\bibitem[Condon et al.(2012)]{Condon12} Condon J.~J., Cotton W.~D., Fomalont E.~B., et al.\ 2012, ApJ, 758, 23

\bibitem[Dewdney  et al.(2013)]{Dewdney13}{Dewdney P., Turner W., Millenaar R., McCool R., Lazio J., Cornwell T.\ 2013, ``SKA1 System Baseline Design", Document number SKA-TEL-SKO-DD-001 Revision 1}

\bibitem[Ferrari et al.(2015)]{ferrari15}{Ferrari C., et al.\ 2015, "Non-thermal emission from galaxy clusters: feasibility study with SKA1", in proceedings of  "Advancing Astrophysics with the Square Kilometre Array",  \pos{PoS(AASKA14)075}}

\bibitem[Gitti et al.(2015)]{gitti15}{Gitti M., et al.\ 2015, "The SKA view of cool-core clusters: evolution of radio mini-halos and AGN feedback", in proceedings of  "Advancing Astrophysics with the Square Kilometre Array",  \pos{PoS(AASKA14)076}} 

\bibitem[Grainge et al.(2015)]{grainge15}{Grainge K., et al.\ 2015, "Galaxy Cluster Science with the Sunyaev-Zel'dovich Effect", in proceedings of  "Advancing Astrophysics with the Square Kilometre Array",  \pos{PoS(AASKA14)170}} 


\bibitem[Jarvis et al.(2015)]{jarvis15}{Jarvis M.~J., et al.\ 2015, "The star-formation history of the Universe with the SKA", in proceedings of  "Advancing Astrophysics with the Square Kilometre Array",  \pos{PoS(AASKA14)068}} 

\bibitem[Kapinska et al.(2015)]{kapinska15}{Kapinska A.~D., et al.\ 2015, "Unravelling lifecycles \& physics of radio-loud AGN in the SKA era", in proceedings of  "Advancing Astrophysics with the Square Kilometre Array",  \pos{PoS(AASKA14)173}} 

\bibitem[Makhathini et al.(2015)]{makhathini15}{Makhathini S., Smirnov O.~M., Jarvis M.~J. and Heywood I.\ 2015, "Morphological classification of radio sources for galaxy evolution and cosmology", in proceedings of  "Advancing Astrophysics with the Square Kilometre Array",  \pos{PoS(AASKA14)81}}

\bibitem[Mancuso et al.(2015)]{mancuso15}{Mancuso C., et al.\ 2015, "Radio Observations of Star Forming Galaxies in the SKA era", in proceedings of  "Advancing Astrophysics with the Square Kilometre Array",  \pos{PoS(AASKA14)82}} 

\bibitem[McAlpine et al.(2015)]{mcalpine15}{McAlpine K., et al.\ 2015, "The Interplay between SF and AGN Activity, and its role in Galaxy Evolution", in proceedings of  "Advancing Astrophysics with the Square Kilometre Array",  \pos{PoS(AASKA14)83}} 

\bibitem[McKean et al.(2015)]{mckean15}{McKean J.~P., et al.\ 2015, "Strong gravitational lensing with the SKA", in proceedings of  "Advancing Astrophysics with the Square Kilometre Array",  \pos{PoS(AASKA14)84}}  

\bibitem[Murphy et al.(2015)]{murphy15}{Murphy E.~J., et al.\ 2015, "The Astrophysics of Star Formation Across Cosmic Time at $\gtrsim 10$ GHz with the Square Kilometer Array", in proceedings of  "Advancing Astrophysics with the Square Kilometre Array",  \pos{PoS(AASKA14)85}}  

\bibitem[Nelson et al.(2013)]{Nelson13}{Nelson E.~J., van Dokkum P.~G., Momcheva I., et al.\ 2013, ApJ, 763, L16}

\bibitem[Norris et al.(2015)]{norris15}{Norris R.~P., et al.\ 2015, "The SKA Mid-frequency All-sky Continuum Survey: Discovering the unexpected and transforming radio-astronomy", in proceedings of  "Advancing Astrophysics with the Square Kilometre Array",  \pos{PoS(AASKA14)86}}  

\bibitem[Orienti et al.(2015)]{orienti15}{Orienti M., D'Ammando F., Giroletti M., Giovannini G. and Panessa F.\ 2015, "The physics of the radio emission in the quiet side of the AGN population with the SKA", in proceedings of  "Advancing Astrophysics with the Square Kilometre Array",  \pos{PoS(AASKA14)87}}   

\bibitem[Smolcic et al.(2015)]{smolcic15}{Smolcic V., et al.\ 2015, "Exploring AGN Activity over Cosmic Time with the SKA", in proceedings of  "Advancing Astrophysics with the Square Kilometre Array",  \pos{PoS(AASKA14)69}}  

\bibitem[Umana et al.(2015)]{umana15}{Umana G., et al.\ 2015, "The impact of SKA on Galactic Radioastronomy: continuum observations", in proceedings of  "Advancing Astrophysics with the Square Kilometre Array",  \pos{PoS(AASKA14)118}}   

\bibitem[Vazza et al.(2015)]{vazza15}{Vazza F., et al.\ 2015, "Filaments of the radio cosmic web: opportunities and challenges for SKA", in proceedings of  "Advancing Astrophysics with the Square Kilometre Array",  \pos{PoS(AASKA14)97}}   

\bibitem[Wolter et al.(2015)]{wolter15}{Wolter A., et al.\ 2015, "Radio investigation of Ultra-Luminous X-ray Sources in the SKA Era", in proceedings of  "Advancing Astrophysics with the Square Kilometre Array",  \pos{PoS(AASKA14)91}}   

\bibitem[Zwart et al.(2015)]{zwart15}{Zwart J., et al.\ 2015, "Astronomy below the Survey Threshold", in proceedings of  "Advancing Astrophysics with the Square Kilometre Array",  \pos{PoS(AASKA14)172}}

\end{thebibliography}
\end{document}